\documentclass{article}

\usepackage[dvips]{graphicx,color} 
\usepackage{dcolumn}
\usepackage{bm}
\usepackage{subfigure} 

\begin{document}


\title{Heterogeneous nucleation on a completely wettable substrate}

\author{Masao Iwamatsu\footnote{e-mail: iwamatsu@ph.ns.tcu.ac.jp, tel:+81-3-5705-1571, fax:+81-3-5707-2222} \\
\\
Department of Physics, Faculty of Liberal Arts and Sciences\\
Tokyo City University\\
Setagaya-ku, Tokyo 158-8557, Japan
}

\date
{
}

\maketitle

\begin{abstract}
It is widely believed that heterogeneous nucleation occurs without an activation process when the surface is completely wettable.   In this report, we review our previous work [J.Chem.Phys {\bf 134}, 234709 (2011)] to show that the critical nucleus (droplet) can exist and the activation process may be observable. In fact, a critical nucleus and a free energy barrier always exist if the surface potential or the disjoining pressure allows for the first-order pre-wetting transition on a completely wettable substrate where the contact angle is zero. Furthermore, the critical nucleus changes character from the critical nucleus of surface phase transition below bulk coexistence (undersaturation) to the critical nucleus of bulk heterogeneous nucleation above the coexistence (oversaturation) when it crosses the coexistence. In this paper the morphology and work of formation of a critical nucleus on a completely-wettable substrate are re-examined to point out the possibility of observing a critical droplet on a completely wettable substrate. 

\end{abstract}


\section*{\label{sec:sec1}1. Introduction}

The nucleation which occurs within the bulk metastable material is called {\it homogeneous nucleation}~\cite{Kelton10,Oxtoby88}.  However, it is frequently assisted by the presence of a surface.  Nucleation in this case is called {\it heterogeneous nucleation}~\cite{Kelton10,Oxtoby88,Turnbull49,Fletcher58,Gretz66,Scheludko81,Joanny86,Lazaridis93,Winter09}.  In the case of vapor to liquid nucleation, for example, a liquid droplet of semi-spherical shape will be formed on the substrate which attracts liquid within the {\it oversaturated} vapor.  The shape of this critical nucleus of the liquid is characterized by the apparent contact angle $\theta_{a}$ shown in Fig.~1.  

The nucleation is characterized by the nucleation rate $J$, which is the number of critical nuclei formed per unit time per unit volume.  Usually it is written in Arrhenius activation form
\begin{equation}
J = A \exp\left(-\frac{W}{k_{\rm B}T}\right)
\label{Eq:1x}
\end{equation}
where $A$ is a kinetic pre-exponential factor which is believed to be weakly dependent on temperature $T$, $k_{\rm B}$ is the Boltzmann constant, and $W$ is the reversible work of formation of critical nucleus

According to classical nucleation theory (CNT)~\cite{Turnbull49}, the work of formation (free energy barrier) of heterogeneous nucleation $W_{\rm hetero}$ is expressed using the apparent contact angle $\theta_{a}$ and the free energy barrier $W_{\rm homo}$ of the homogeneous nucleation as 
\begin{equation}
W_{\rm hetero} = W_{\rm homo}f\left(\theta_{a}\right),
\label{Eq:2x}
\end{equation}
where
\begin{equation}
f\left(\theta_{a}\right)=\left(\theta_{a}-\cos\theta_{a}\sin\theta_{a}\right)/\pi
\label{Eq:3x}
\end{equation}
for the two-dimensional semi-cylindrical nucleus, and 
\begin{equation}
f\left(\theta_{a}\right)=\left(1-\cos\theta_{a}\right)^{2}\left(2+\cos\theta_{a}\right)/4.
\label{Eq:4x}
\end{equation}
for the three-dimensional axi-symmetric semi-spherical nucleus.  The contact angle $\theta_{a}$ characterizes the interaction between the liquid and substrate.  When $\pi>\theta_{a}>0$ the liquid is said to incompletely wet the substrate.  When $\theta_{a}=0$, the liquid is said to completely wet the substrate. 
Since $0< f\left(\theta_{a}\right)<1$, the presence of the substrate enhances the nucleation rate (Eq.~(\ref{Eq:1x})) when the liquid incompletely wets the substrate ($\pi>\theta_{a}>0$) as $W_{\rm hetero}<W_{\rm homo}$.  In a completely wettable substrate where the liquid completely wets the substrate ($\theta_{a}=0$), CNT predicts that no nucleation barrier exists ($W_{\rm hetero}=0$) as $f\left(\theta_{a}=0\right)=0$.

\begin{figure}[htbp]
\begin{center}
\includegraphics[width=1.0\linewidth]{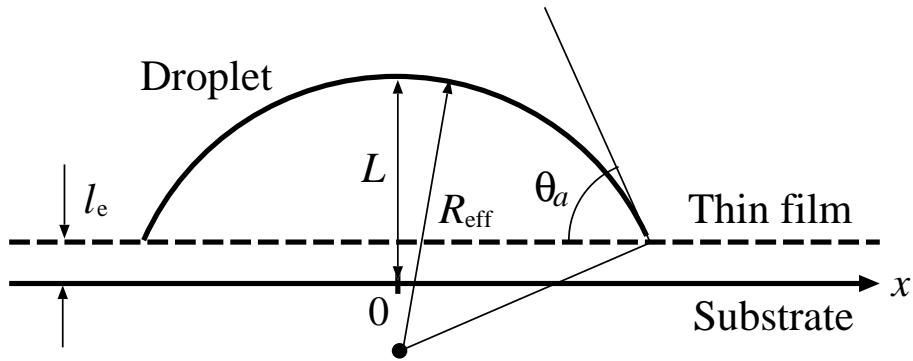}
\caption{
An ideal critical nucleus of heterogeneous nucleation on a substrate.  The apparent contact angle $\theta_{a}$ is defined as the angle at the intersection of the cylindrical or spherical droplet surface with an effective radius $R_{\rm eff}$ and height $L$ from the substrate with a wetting layer of thickness $l_{e}$.
}
\label{fig:1}
\end{center}
\end{figure}

This heterogeneous nucleation of a new bulk phase on a flat substrate can be associated with the surface phase transition called wetting transition as both phase transitions can occur on the same surface~\cite{Talanquer96}.  In fact, there exists a critical nucleus and a free energy barrier if the first-order so-called prewetting transition~\cite{Dietrich88,Bonn01,Bonn09} can occur on a completely-wettable substrate.  This transition occurs in the undersaturated vapor and separates states of thin and thick liquid films on a completely-wettable substrate with the zero contact angle. The shape of the critical nucleus is flat and pancake-like~\cite{Joanny86b}, and cannot be characterized simply by a contact angle.  Therefore, in contrast to the prediction of CNT, the critical nucleus and the free energy barrier for heterogeneous nucleation can exist even if the substrate is completely-wettable and the contact angle $\theta_{a}=0$ (Eq.~(\ref{Eq:2x})). By varying the chemical potential or the vapor pressure from undersaturation to oversaturation, the droplet changes character from the critical nucleus of the prewetting surface phase transition to that of the vapor to liquid bulk phase transition.  Recently, Sear~\cite{Sear08} has shown by a direct numerical calculation of the nucleation rate that the nucleus "does not notice" this change when it crosses the coexistence. 

In this paper, we summarize the previous study~\cite{Iwamatsu2011} and the morphology and the work of formation of a critical nucleus rather than the nucleation rate~\cite{Sear08} on a completely-wettable flat and spherical substrates are re-examined across the coexistence using the interface-displacement model (IDM).  This model has been successfully used for studying wetting phenomena~\cite{Dietrich88,Bonn01,Bonn09}, line tensions~\cite{Indekeu92,Dobbs93} and even layering transitions~\cite{Iwamatsu98}.   In fact, various properties of critical nucleus of prewetting transition rather than that of the bulk critical nucleus have already been studied~\cite{Bausch92,Bausch93,Blossey95,Bausch96,Nakanishi82,Blockhuis95}.  However other authors paid most attention to the nucleation of wetting transition in the undersaturated vapor {\it below} the bulk coexistence.  In fact, the heterogeneous nucleation of bulk phase transition occurs in the oversaturated vapor {\it above} the bulk coexistence.  Therefore, we will pay attention to those properties which are relevant to the bulk heterogeneous nucleation.  In this case only a few authors such as Talanquer and Oxtoby~\cite{Talanquer96} and Sear~\cite{Sear08} have considered such properties.

\section*{2. Interface-displacement model}

Within the interface displacement model (IDM) in $d$-dimensional space the free energy $\Omega$ of a liquid film of local thickness $l\left({\bf x}\right)$ is given by~\cite{Wyart91,Bauer99,Yeh99,Dobbs99,Starov09,Bausch92,Bausch93,Blossey95}
\begin{equation}
\Omega\left[l\right]=\int\left[\gamma \left(\left(1+ \left(\nabla l\right)^{2}\right)^{1/2}-1\right)+V\left( l\right)-\mu l\right]d^{d-1}x,
\label{Eq:5x}
\end{equation}
where $\gamma$ is the liquid-vapor surface tension, $V\left(l\right)$ is the effective interface potential~\cite{Derjaguin87,Dietrich88,Bonn09,Dietrich91} from the substrate,  and $\mu$ denotes the deviation of the chemical potential from liquid-vapor coexistence such that $\mu=0$ is the bulk coexistence and for $\mu>0$ the vapor is oversaturated. In Eq.~(\ref{Eq:5x}) we keep the nonlinear dependence on $\nabla l$. The morphology and the apparent contact angle $\theta_{a}$ of the critical nucleus of heterogeneous nucleation will be determined by minimizing Eq.~(\ref{Eq:5x}).  

\begin{figure}[htbp]
\begin{center}
\includegraphics[width=1.0\linewidth]{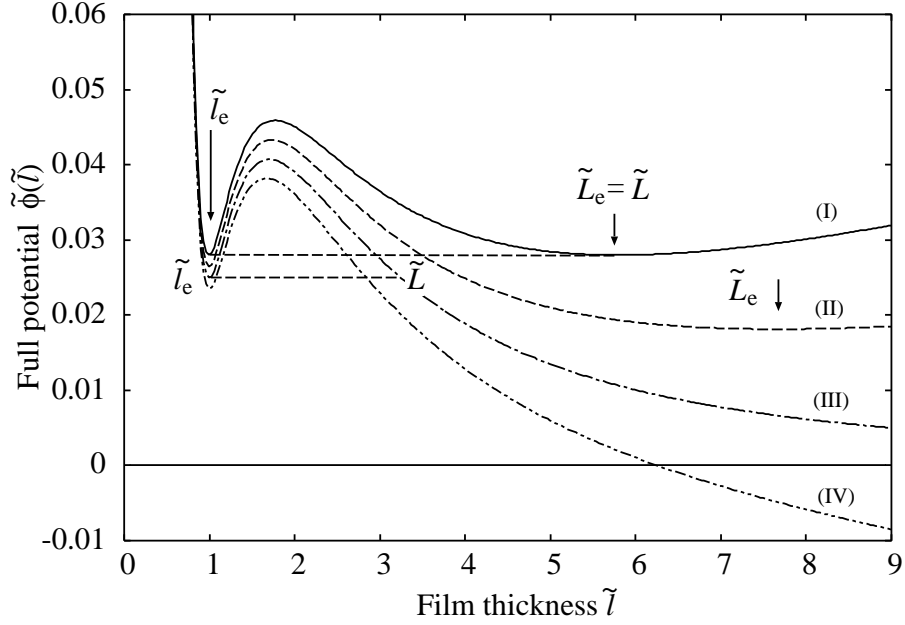}
\caption{
Typical shapes of the full potential $\phi\left(l\right)$ as a function of the film thickness $l$. Reduced full potential $\tilde{\phi}(\tilde{l})$ (Eq.(\ref{Eq:23x})), which will be defined in section 3, as a function of the reduced film thickness $\tilde{l}$ (Eq.~(\ref{Eq:24x})) of a completely-wettable substrate for various reduced chemical potential $\tilde{\mu}$ (Eq.~(\ref{Eq:24x})). (I) at the prewetting line $\mu=\mu_{\rm p}$ ($\tilde{\mu}_{\rm p}=-0.00299$) where a thin film and thick film coexist, (II) below the bulk coexistence and above the prewetting line $0>\mu>\mu_{\rm p}$ ($\tilde{\mu}=-0.0015$) where the thick film is stable, (III) at the bulk coexistence $\mu=0$ ($\tilde{\mu}=0.0000$) where the thick film becomes infinitely thick, (IV) above the bulk coexistence $\mu>0$ ($\tilde{\mu}=+0.0015$) where the vapor is oversaturated and the bulk liquid phase becomes stable. The $d=2$-dimensional droplet height $L$ is determined from the energy conservation law (\ref{Eq:18x}). 
}
\label{fig:2}
\end{center}
\end{figure}

Sometimes it is useful to consider the full potential $\phi(l)$, which includes the chemical potential $\mu$ defined by
\begin{equation}
\phi(l)=V(l)-\mu l.
\label{Eq:6x}
\end{equation}
instead of the effective interface potential $V\left(l\right)$.  This full potential $\phi(l)$ depends not only on the chemical potential $\mu$ but also on the temperature $T$.  Figure \ref{fig:2} shows typical shapes of the full potential $\phi(l)$ of a completely-wettable substrate.  The full potential $\phi(l)$ exhibits double-well shape typical of the first-order surface phase transition, and its two minimums at $l_{e}$ and at $L_{e}$ correspond to the metastable thin and the stable thick wetting films.  Figure~\ref{fig:2} indicates that the thick wetting film becomes infinitely thick ($L_{e}=\infty$) when $\mu\ge 0$.  

\begin{figure}[htbp]
\begin{center}
\includegraphics[width=0.9\linewidth]{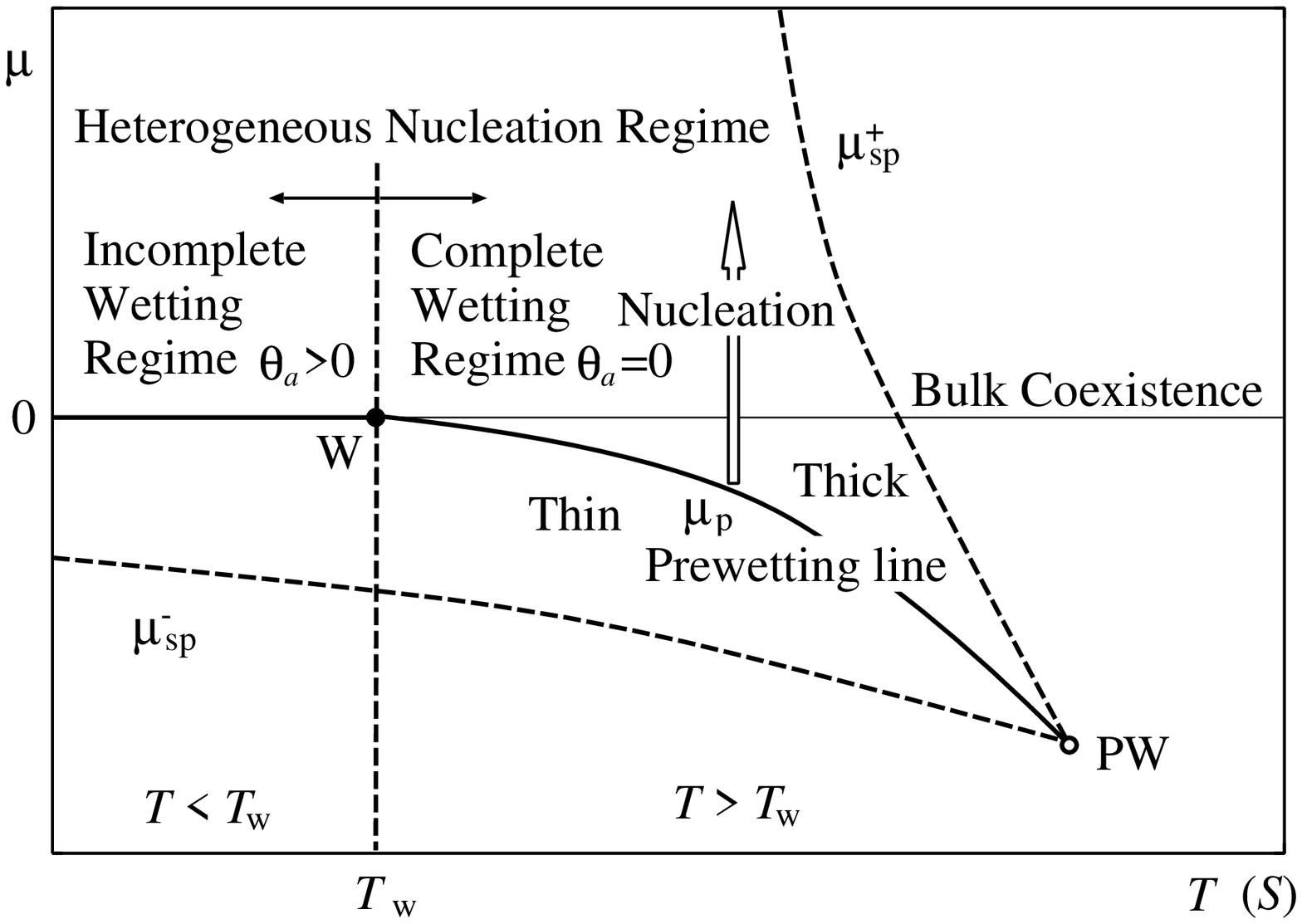}
\caption{
Schematic surface phase diagram in the spreading coefficient $S$ or the temperature $T$ and chemical potential $\mu$ plane.  The horizontal line $\mu=0$ corresponds to the bulk coexistence line, on which at the wetting transition point W with $S=0$ ($T=T_{\rm w}$) a first-order wetting transition occurs.  The prewetting transition line $\mu_{\rm p}$ starts at W and terminates at the prewetting critical point PW.  Above this prewetting line and below the bulk coexistence ($0>\mu>\mu_{\rm p}$), the stable state is a thick film.  The thickness diverges at the bulk coexistence and this infinitely thick film becomes bulk liquid phase above the coexistence ($\mu\ge 0$).  A large arrow indicates the route of the isothermal heterogeneous nucleation on a completely-wettable substrate.}
\label{fig:3}
\end{center}
\end{figure}

The schematic surface phase diagram is shown in Fig.~\ref{fig:3} in the $T$-$\mu$ (temperature-chemical potential) plane~\cite{Bonn01,Bonn09,Indekeu92,Dobbs93,Bausch92,Blossey95}.  Since the local minimum $V\left(l_{e}\right)$ at $l_{e}$ is related to the spreading coefficient $S$ and the temperature $T$ through~\cite{Bausch92,Blossey95}
\begin{equation}
\phi\left(l_{e}\right)=S\propto T-T_{\rm w},
\label{Eq:7x}
\end{equation}
where $T_{\rm w}$ is the wetting temperature, and $S$ is defined by
\begin{equation}
S=\gamma_{\rm sv}-\gamma_{\rm sl}-\gamma,
\label{Eq:8x}
\end{equation}
where $\gamma_{\rm sv}$ and $\gamma_{\rm sl}$ are the substrate-vapor and the substrate-liquid interfacial tensions, the first-order wetting transition from the incomplete wetting of a thin liquid film with thickness $l_{e}$ ($\phi\left(l_{e}\right)<0=\phi\left(l=\infty\right)$) to the complete wetting of an infinite thickness with $l=\infty$ ($\phi\left(l_{e}\right)>0=\phi\left(l=\infty\right)$) will occur at the wetting transition point W at $T=T_{\rm w}$ (wetting temperature) along $\mu=0^{-}$ in Fig.~\ref{fig:3}. From the Young's formula
\begin{equation}
\gamma\cos\theta_{a}=\gamma_{\rm sv}-\gamma_{\rm sl}
\label{Eq:9x}
\end{equation}
we have
\begin{equation}
\cos\theta_{a}=1+\frac{S}{\gamma}.
\label{Eq:10x}
\end{equation}
Therefore the apparent contact angle vanishes ($\theta_{a}=0$) in the complete-wetting regime with $S\ge 0$, and so does the free-energy barrier Eq.~(\ref{Eq:2x}).

Next, we will discuss the surface phase diagram in the complete-wetting regime ($S>0, \;\;T>T_{\rm w}$).  In this regime, the prewetting transition $\mu_{\rm p}(T)$($<0$) line appears below the bulk coexistence line $\mu=0$ (Fig.~\ref{fig:3}). At the prewetting chemical potential $\mu=\mu_{\rm p}$, the full potential $\phi(l)$ has double-minimum shape with the same depth shown as the curve (I) in Fig.~\ref{fig:2}, which satisfies
\begin{equation}
\phi\left(l_{e}\right) = \phi\left(L_{e}\right),
\label{Eq:11x}
\end{equation}
and
\begin{equation}
\left.\frac{d\phi}{dl}\right|_{l_{e}} = \left.\frac{d\phi}{dl}\right|_{L_{e}}=0.
\label{Eq:12x}
\end{equation}
Then the thin (thickness $l_{e}$) and the thick (thickness $L_{e}$) wetting films can coexist at $\mu=\mu_{\rm p}<0$, which plays the role of bulk coexistence $\mu=0$ for the heterogeneous nucleation in the complete-wetting regime ($T>T_{\rm w}$). When the chemical potential is increased ($0>\mu>\mu_{\rm p}$) along the arrow indicated in Fig.~\ref{fig:3}, the thick film with thickness $L_{e}$ becomes stable and the thin film with $l_{e}$ becomes metastable above the prewetting line  ($\phi\left(L_{e}\right)<\phi\left(l_{e}\right)$, curve (II) in Fig.~\ref{fig:2}). At and above the bulk coexistence $\mu\geq 0$ along the arrow in Fig.~\ref{fig:3}, the thickness of the stable thick film diverges (curves (III) and (IV) in Fig.~\ref{fig:2}).  Finally, the  metastable thin film loses stability at the upper-spinodal  $\mu=\mu_{\rm sp}^{+}$ shown in Fig.~\ref{fig:3}.

\section*{3. Nucleation on a completely-wettable flat substrate}

\subsection*{3.1 $d=2$-dimensional nucleus}
Based on the phase diagram shown in Fig.~\ref{fig:3} and the morphology of the potential $\phi(l)$ shown in Fig.~\ref{fig:2}, we can discuss the morphology and the work of formation of the critical nucleus on a completely-wettable substrate where the apparent contact angle vanishes ($\theta_{a}=0$).  

For a $d=2$-dimensional cylindrical droplet, the Euler-Lagrange equation for the free energy given by Eq.~(\ref{Eq:5x}) is simplified to~\cite{Bauer99,Yeh99,Dobbs99,Starov09}
\begin{equation}
\frac{\gamma l_{xx}}{\left(1+l_{x}^{2}\right)^{3/2}}=\frac{d\phi}{dl},
\label{Eq:14x}
\end{equation}
where $l_{xx}=d^{2}l/dx^{2}$.  Equation (\ref{Eq:14x}) could be considered as a kind of equation of motion for a classical particle moving in a potential $-\phi(l)$.

Equation (\ref{Eq:14x}) can be integrated once to give
\begin{equation}
\frac{-\gamma}{\left(1+l_{x}^{2}\right)^{1/2}}=-\gamma\cos\theta\left( l\right)=V\left(l\right)-\mu l+C,
\label{Eq:15x}
\end{equation} 
where $C$ is the integration constant and $\cos\theta\left( l\right)$ is the cosine of the angle $\theta(l)$ made between the tangential line of the liquid-vapor surface at the height $l(x)$ and the substrate~\cite{Yeh99}.  

Near the substrate, since the liquid-vapor interface will smoothly connect to the surrounding thin liquid film of thickness $l=l_{e}$ with $\theta\left(l_{e}\right)=0$, the integration constant $C$ in Eq.~(\ref{Eq:15x}) will be given by $C=-\gamma-V\left( l_{e}\right)+\mu l_{e}$, and the liquid-vapor interface will be determined from
\begin{equation}
\frac{-\gamma}{\left(1+l_{x}^{2}\right)^{1/2}}=\left(V\left(l\right)-V\left( l_{e}\right)\right)-\mu\left(l-l_{e}\right)-\gamma.
\label{Eq:16x}
\end{equation} 
Similarly, at the top of the droplet with a height $l=L$, we have again $\theta\left(L\right)=0$ at $l=L$ (Fig.~\ref{fig:2}), and the liquid-vapor interface will be determined from an equation similar to Eq.~(\ref{Eq:16x}) with $l_{e}$ replaced by $L$:
\begin{equation}
\frac{-\gamma}{\left(1+l_{x}^{2}\right)^{1/2}}=\left(V\left(l\right)-V\left( L\right)\right)-\mu\left(l-L\right)-\gamma.
\label{Eq:17x}
\end{equation}
Since Eqs.~(\ref{Eq:16x}) and (\ref{Eq:17x}) must be identical, we have
\begin{equation}
V\left(l_{e}\right)-\mu l_{e} = V\left( L\right)-\mu L,\;\;\;\mbox{or}\;\;\;\phi\left(l_{e}\right)=\phi\left(L\right),
\label{Eq:18x}
\end{equation}
which is similar to the energy conservation law for a classical particle whose pseudo-equation of motion is given by Eq.~(\ref{Eq:14x}) moving in a potential $-\phi(l)$.  Then, the height $L$ of the $d=2$-dimensional cylindrical droplet can be determined from Eq.~(\ref{Eq:18x})~\cite{Dobbs93}.  Equations (\ref{Eq:17x}) and (\ref{Eq:18x}) can also be transformed into a single expression using the disjoining pressure~\cite{Derjaguin87,Boinovich2011}.

Therefore a cylindrical droplet can exist even on a completely-wettable substrate with $\theta_{a}=0$.  Also, it exists even in an {\it undersaturated} vapor below the bulk coexistence and above the prewetting line  ($0>\mu>\mu_{\rm p}$) in the thick film phase (Fig.~\ref{fig:3}). In this thick film phase, the droplet is not the critical nucleus of the bulk phase transition, but is in fact the {\it critical nucleus} of the thin-thick {\it surface phase transition}. This critical nucleus is expected to transform continuously into the {\it critical nucleus} of the {\it bulk phase transition} above the bulk coexistence ($\mu_{\rm sp}^{+}>\mu>0$) when the chemical potential crosses the coexistence $\mu=0$.  In contrast, it must be noted that the critical nucleus, and therefore, a spherical droplet, can exist on an incompletely-wettable substrate only in an {\it oversaturated} vapor~\cite{Boinovich2009}.

In order to study the film thickness $l_{e}$, $L_{e}$ and the droplet height $L$ more quantitatively, a model interface potential given by
\begin{equation}
V\left(l\right)=V_{0}\left(\frac{1}{2}\left(\frac{l_{0}}{l}\right)^{2} - \frac{1+b}{3}\left(\frac{l_{0}}{l}\right)^{3} + \frac{b}{4}\left(\frac{l_{0}}{l}\right)^{4}\right)
\label{Eq:19x}
\end{equation}
is used, where $l_{0}$ is a typical thickness of the thin film, and $V_{0}>0$ plays the role of the so-called Hamaker constant~\cite{Israelachvili92,Iwamatsu98} $A_{\rm slv}$ of the substrate-liquid-vapor system through
\begin{equation}
\frac{V_{0}l_{0}^{2}}{2}=\frac{A_{\rm slv}}{12\pi}.
\label{Eq:20x}
\end{equation}
Equation (\ref{Eq:19x}) is derived from the long-ranged potential of the form $v\left(r\right)\sim 1/r^{n}, n=6, 7, 8$, but a more general form will be required to study real situation~\cite{Boinovich2011}.  Since, we are most interested in conceptual problem of the possibility of droplet on a completely-wettable substrate, we will use the simplest form in Eq.~(\ref{Eq:19x}).

The parameter $b$ plays the role of the temperature that controls the transition from incomplete- to complete-wetting.  Since the two minima of Eq.~(\ref{Eq:19x}) are located at $l/l_{0}=1$ and $l/l_{0}=\infty$, and the spreading coefficient $S$ is related to the parameter $b$ through
\begin{equation}
S=V\left(l_{e}\right)=V_{0}\frac{2-b}{12},
\label{Eq:21x}
\end{equation}
a complete wetting with $S>0$ is achieved when $b<2$.  We use the potential parameter $b=1.7$ in the completely-wetting regime, which has already been used in Fig.~\ref{fig:2}.  

\begin{figure}[htbp]
\begin{center}
\includegraphics[width=1.0\linewidth]{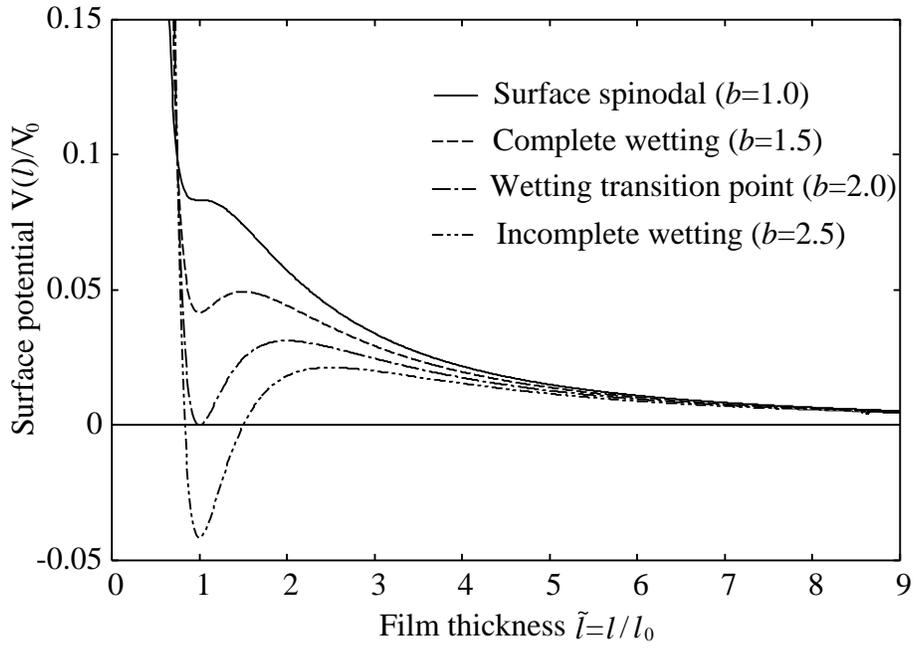}
\caption{
The reduced surface potential $V(l)/V_{0}$ for various $b$ which corresponds to complete to incomplete wetting regimes. }
\label{fig:3-2}
\end{center}
\end{figure}

From Eq.~(\ref{Eq:19x}), the full potential (Eq.~(\ref{Eq:6x})) can be written as
\begin{equation}
\phi\left(l\right) = V_{0} \tilde{\phi}\left(\tilde{l}\right)
\label{Eq:22x}
\end{equation}
using non-dimensional reduced potential $\tilde{\phi}$ defined by
\begin{equation}
\tilde{\phi}\left(\tilde{l}\right)=\frac{1}{2\tilde{l}^{2}}-\frac{1+b}{3\tilde{l}^{3}}+\frac{b}{4\tilde{l}^{4}}-\tilde{\mu}\tilde{l},
\label{Eq:23x}
\end{equation}
and
\begin{equation}
\tilde{\mu}=\frac{\mu l_{0}}{V_{0}},\;\;\;\tilde{l} = \frac{l}{l_{0}}.
\label{Eq:24x}
\end{equation}
Figure \ref{fig:3-2} shows the typical shapes of surface potential $V$ (reduced potential $\tilde{\phi}$ at $\mu=0$) for various $b$ when $\mu=0$ (liquid-vapor coexistence) in complete and incomplete wetting regimes.  In Fig.~\ref{fig:2} we have already shown the reduced potential $\tilde{\phi}$ in the complete-wetting regime for various reduced chemical potentials $\tilde{\mu}$ below the bulk coexistence $\tilde{\mu}=0$ and above the prewetting $\tilde{\mu}_{\rm p}$.  We set $b=1.7$ for which the prewetting chemical potential is given by $\tilde{\mu}_{\rm p}=-0.00299$.  

\begin{figure}[htbp]
\begin{center}
\includegraphics[width=1.0\linewidth]{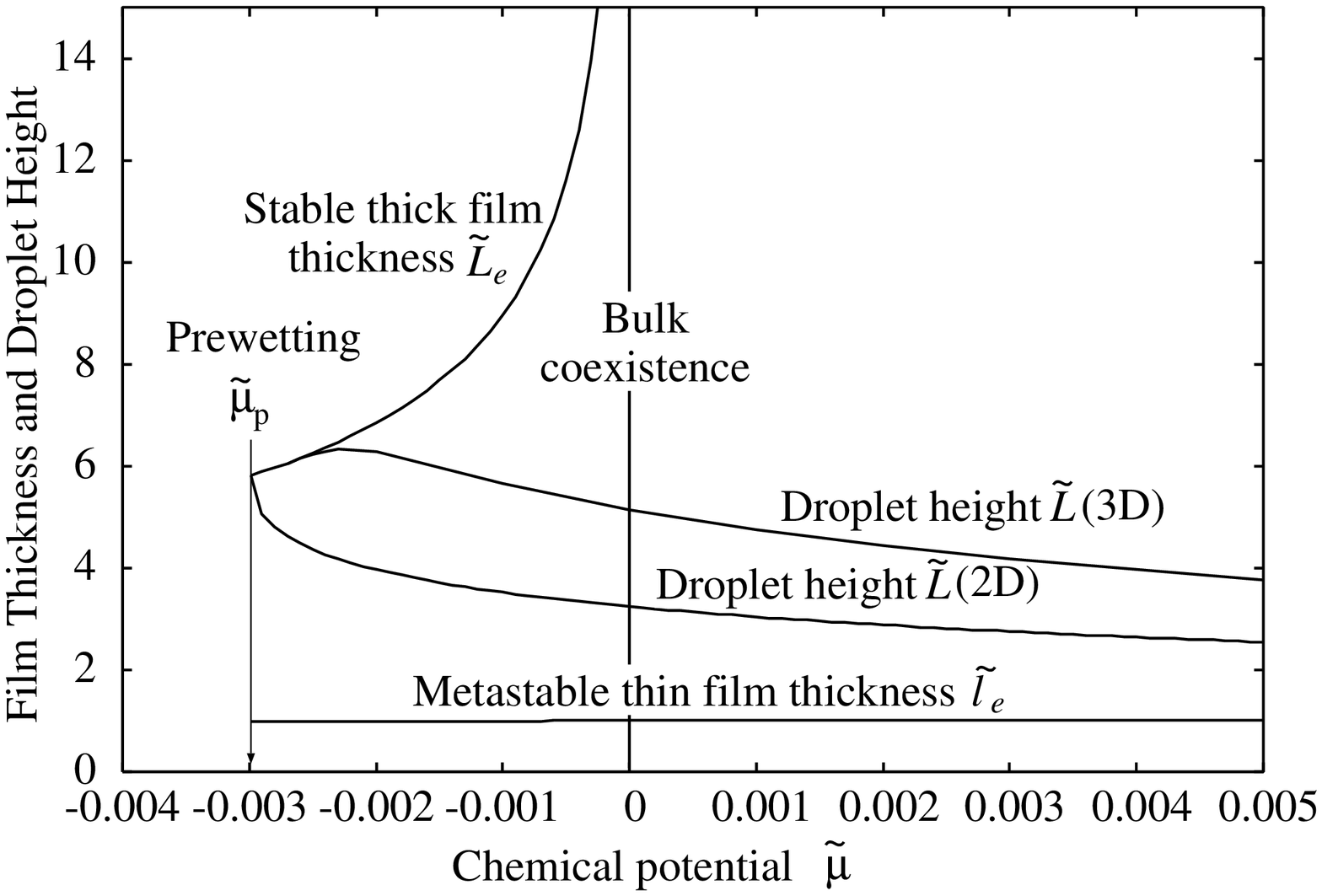}
\caption{
The stable thick film thickness $\tilde{L}_{e}=L/l_{0}$ and the metastable thin film thickness $\tilde{l}_{e}=l_{e}/l_{0}$ calculated from the model interface potential Eq.~(\ref{Eq:23x}).  The droplet height $\tilde{L}=L/l_{0}$ of the $d=2$-dimensional (2D) cylindrical nucleus determined from Eq.~(\ref{Eq:18x}) and that of the $d=3$-dimensional (3D) semi-spherical nucleus that will be determined from the boundary value problem of the Euler-Lagrange equation (\ref{Eq:47x}) in the next subsection are also shown.  }
\label{fig:4}
\end{center}
\end{figure}

Figure~\ref{fig:4} shows the reduced stable thick film thickness $\tilde{L}_{e}=L_{e}/l_{0}$, the metastable thin film thickness $\tilde{l}_{e}=l_{e}/l_{0}$, and the droplet height $\tilde{L}=L/l_{0}$ of the $d=2$-dimensional cylindrical nucleus determined from Eq.~(\ref{Eq:18x}) in the complete wetting regime with $b=1.7$. The droplet height $L$ must be equal to the thick film thickness $L_{e}$ and remain finite at the prewetting line ($\mu=\mu_{\rm p}$).  We observe that all physical quantities change continuously at the bulk coexistence ($\mu=0$) when the chemical potential $\mu$ is increased from negative ($\mu<0$, undersaturation) to positive ($\mu>0$, oversaturation). 

Using the expansion $V\left(l\right)-V\left( L\right)\simeq (l-L)\left.(dV/dl)\right|_{l=L}$, we have from Eq.~(\ref{Eq:17x})
\begin{equation}
\frac{-\gamma}{\left(1+l_{x}^{2}\right)^{1/2}}=\left(\left.\frac{dV}{dl}\right|_{l=L}-\mu\right)\left(l-L\right)-\gamma
\label{Eq:25x}
\end{equation}
near the top of the droplet $l\simeq L$.  If the effective chemical potential defined by
\begin{equation}
\mu_{\rm eff}=-\left.\frac{dV}{dl}\right|_{l=L}+\mu =-\left.\frac{d\phi}{dl}\right|_{l=L}
\label{Eq:26x}
\end{equation}
at the droplet top with height $L$ is positive, then the solution of Eq.~(\ref{Eq:17x}) near the top of the droplet is given by a semi-circular shape (Fig.~\ref{fig:1}) 
\begin{equation}
l=\sqrt{R_{\rm eff}^2-x^2}-\left(R_{\rm eff}-L\right)
\label{Eq:28x}
\end{equation}
where $R_{\rm eff}-L$ is the shift of the base of circular interface.  The effective radius $R_{\rm eff}$ is given by the Kelvin-Laplace formula~\cite{Joanny86}
\begin{equation}
R_{\rm eff}=\frac{\gamma}{\mu_{\rm eff}}.
\label{Eq:29x}
\end{equation}

At the prewetting $\mu=\mu_{\rm p}$, the effective chemical potential $\mu_{\rm eff}$ vanishes
\begin{equation}
\mu_{\rm eff}=-\left.\frac{dV}{dl}\right|_{l=L}+\mu_{\rm p}=0
\label{Eq:30x}
\end{equation}
from Eq.~(\ref{Eq:12x}) since $L=L_{e}$.  Hence, the prewetting line $\mu=\mu_{\rm p}$ acts as an effective or a shifted bulk coexistence with $\mu_{\rm eff}=0$ for the critical nucleus.  In fact, Blossey~\cite{Blossey95} has noted that the $d$-dimensional prewetting line corresponds to the $(d-1)$-dimensional bulk coexistence line. 

From Eqs.~(\ref{Eq:26x}) and (\ref{Eq:30x}), the effective chemical potential $\mu_{\rm eff}$ is approximately written as
\begin{equation}
\mu_{\rm eff}\simeq \left(1-\left(\left.\frac{dL}{d\mu}\right|_{\mu=\mu_{\rm p}}\right)\left(\left.\frac{d^{2}V}{dl^{2}}\right|_{l=L}\right)\right)\left(\mu-\mu_{\rm p}\right)
\label{Eq:31x}
\end{equation}
near the prewetting line $\mu=\mu_{\rm p}$, and the radius $R_{\rm eff}$ [Eq.~(\ref{Eq:29x})] of the curvature at the top of the droplets diverges as
\begin{equation}
R_{\rm eff}=\frac{\gamma}{\mu_{\rm eff}}\propto\frac{\gamma}{\mu-\mu_{\rm p}}
\label{Eq:32x}
\end{equation}
at the prewetting line $\mu=\mu_{\rm p}$.  We note that the effective chemical potential $\mu_{\rm eff}$ and, hence, the effective radius $R_{\rm eff}$ (Eq.~(\ref{Eq:29x})) at the top of the droplet change continuously at the bulk coexistence ($\mu=0$).  Physically, the undersaturated vapor turns to an effectively oversaturated vapor due to the effective interface potential of the substrate which prefers to attract vapor to form a thick wetting film.

The lateral size and the liquid-vapor interface of the droplets can only be studied quantitatively by solving the Euler-Lagrange equation (\ref{Eq:14x}), which can be done using the standard numerical method such as the Runge-Kutta method. To this end, we have to fix the parameter $V_{0}/\gamma$ which can be expressed by using the Hamakar constant $A_{\rm slv}$ in Eq.~(\ref{Eq:20x}) and $A_{\rm lvl}$ by
\begin{equation}
\frac{V_{0}}{\gamma}=-\frac{4D_{0}^{2}A_{\rm slv}}{l_{0}^{2}A_{\rm lvl}},
\label{Eq:33x}
\end{equation}
where an empirical formula~\cite{Israelachvili92}
\begin{equation}
\gamma = \frac{A_{\rm lvl}}{24\pi D_{0}^{2}}
\label{Eq:34x}
\end{equation}
with $D_{0}=0.165$nm is used. Suppose we tentatively set $l_{0}=2D_{0}$, and using the combining relationship~\cite{Israelachvili92,Iwamatsu98}
\begin{eqnarray}
A_{\rm slv} &=& -\sqrt{A_{\rm ll}}\left(\sqrt{A_{\rm ss}}-\sqrt{A_{\rm ll}}\right),
\nonumber \\
A_{\rm lvl} &=& A_{\rm ll},
\label{Eq:35x}
\end{eqnarray}
we have
\begin{equation}
\frac{V_{0}}{\gamma} \sim \frac{A_{\rm slv}}{A_{\rm \rm lvl}} \sim \frac{\sqrt{A_{\rm ss}}-\sqrt{A_{\rm ll}}}{\sqrt{A_{\rm ll}}},
\label{Eq:36x}
\end{equation}
which will be $V_{0}/\gamma\sim 0.1-10$ using typical values of $A_{\rm ss}$ and $A_{\rm ll}$~\cite{Iwamatsu98}. By using the scaled quantities $\tilde{\mu}_{\rm eff}=\mu_{\rm eff}l_{0}/V_{0}$ and $\tilde{R}_{\rm eff}=R_{\rm eff}/l_{0}$, Eq.~(\ref{Eq:29x}) can be written as
\begin{equation}
\tilde{R}_{\rm eff} = \frac{1}{\left(V_{0}/\gamma\right)\tilde{\mu}_{\rm eff}}.
\label{Eq:37x}
\end{equation}
Therefore, the lateral size of nucleus which is roughly determined from $R_{\rm eff}$ is in inverse proportion to $V_{0}/\gamma$.

\begin{figure}[htbp]
\begin{center}
\includegraphics[width=1.0\linewidth]{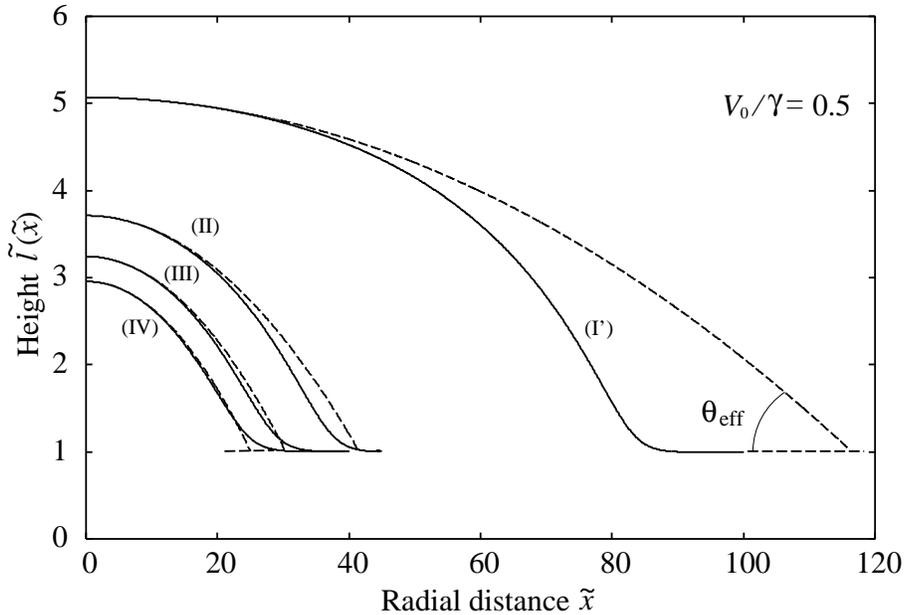}
\caption{
The droplet shape numerically determined from the simplified Euler-Lagrange equation Eq.~(\ref{Eq:14x}) using the Runge-Kutta method (solid curves) compared with the ideal semi-circular shape given by Eq.~(\ref{Eq:28x}) (broken curves) (I') near the prewetting line ($\tilde{\mu}=-0.0029$), (II) below the bulk coexistence and above the prewetting line $0>\mu>\mu_{\rm p}$ ($\tilde{\mu}=-0.0015$), (III) at the bulk coexistence $\mu=0$ ($\tilde{\mu}=0.0000$), (IV) above the bulk coexistence $\mu>0$ ($\tilde{\mu}=+0.0015$).  Only a right half of the droplet is shown.  Note the very flat droplet shape suggested from the scale of the vertical and the horizontal axes.  } 
\label{fig:6}
\end{center}
\end{figure}

Figure \ref{fig:6} compares numerically determined $d=2$-dimensional cylindrical droplet shapes with the ideal semi-circular shapes (Eq.~(\ref{Eq:28x})) with the height $L$ and the effective radius $R_{\rm eff}$ calculated from Eqs.~(\ref{Eq:18x}) and (\ref{Eq:29x}) when $V_{0}/\gamma=0.5$.  The droplet shape deviates significantly from an ideal circular shape and becomes very flat, in particular, below the bulk coexistence $\mu<0$.  This shape is called a pancake~\cite{Joanny86b} as the chemical potential is decreased down to the prewetting line $\mu_{\rm p}$.  The size of the critical pancake is finite even at the prewetting line (Fig.~\ref{fig:6}) even though the effective radius $R_{\rm eff}$ diverges (Eq.~(\ref{Eq:32x})).  

It is possible to define an effective contact angle $\theta_{\rm eff}$ (see Fig.~\ref{fig:6}) of a droplet on a completely-wettable substrate by extrapolating the semi-circular shape (Eq.~(\ref{Eq:28x})) down to the surface of the thin film with thickness $l_{e}$. From the geometrical consideration shown in Fig.~\ref{fig:1}, we find
\begin{equation}
\cos\theta_{\rm eff}=1-\frac{L}{R_{\rm eff}}=1-\frac{\mu_{\rm eff}L}{\gamma}.
\label{Eq:38x}
\end{equation}
from Eq.~(\ref{Eq:29x}).  Therefore, the effective contact angle is very small ($\theta_{\rm eff}\ll 1$) since $l\ll R_{\rm eff}$ and will vanish ($\theta_{\rm eff}\rightarrow 0$) at the prewetting line because $L\rightarrow L_{e}$ and $R_{\rm eff}\rightarrow \infty$ as $\mu\rightarrow \mu_{\rm p}$ from Eq.~(\ref{Eq:32x}).

The work of formation $W$ in Eq.~(\ref{Eq:1x}) can be calculated by inserting the droplet profile in Fig.~\ref{fig:6} obtained from the Euler-Lagrange equation (\ref{Eq:14x}) into Eq.~(\ref{Eq:5x}) with $d=2$ and subtracting the contribution from the wetting film with thickness $l_{e}$:
\begin{eqnarray}
W &=& \int\left[\gamma \left(\left(1+l_{x}^{2}\right)^{1/2}-1\right) \right. \nonumber \\
&+& \left. \left(V\left(l\right)-\mu l\right) - \left(V\left(l_{e}\right)-\mu l_{e}\right) \right]dx,
\label{Eq:39x}
\end{eqnarray}
which can be rewritten using Eq.~(\ref{Eq:16x}) as~\cite{Dobbs93}
\begin{eqnarray}
W &=& \int_{l=l_{e}}^{l=L} \frac{\gamma l_{x}}{\left(1+l_{x}^{2}\right)^{1/2}} dl,
\nonumber \\
&=& \int_{l=l_{e}}^{l=L} \left(2\gamma\Delta\phi\left(l\right)-\Delta\phi\left(l\right)^{2}\right)^{1/2} dl,
\label{Eq:40x}
\end{eqnarray}
where
\begin{equation}
\Delta\phi\left(l\right)=\phi\left(l\right)-\phi\left(l_{e}\right)=\phi\left(l\right)-\phi\left(L\right).
\label{Eq:41x}
\end{equation}
Equation (\ref{Eq:40x}) is also known as the {\it line tension}~\cite{Indekeu92,Dobbs93,Dobbs99}.  Here we interpret this energy as the work of formation of the critical nucleus on a completely-wettable substrate.  Intuitively, the critical nucleus in the complete-wetting regime is approximated by a thin flat disk (cf. Fig.~\ref{fig:6}) and its free energy is given only by the line tension of its perimeter.   

\begin{figure}[htbp]
\begin{center}
\includegraphics[width=1.0\linewidth]{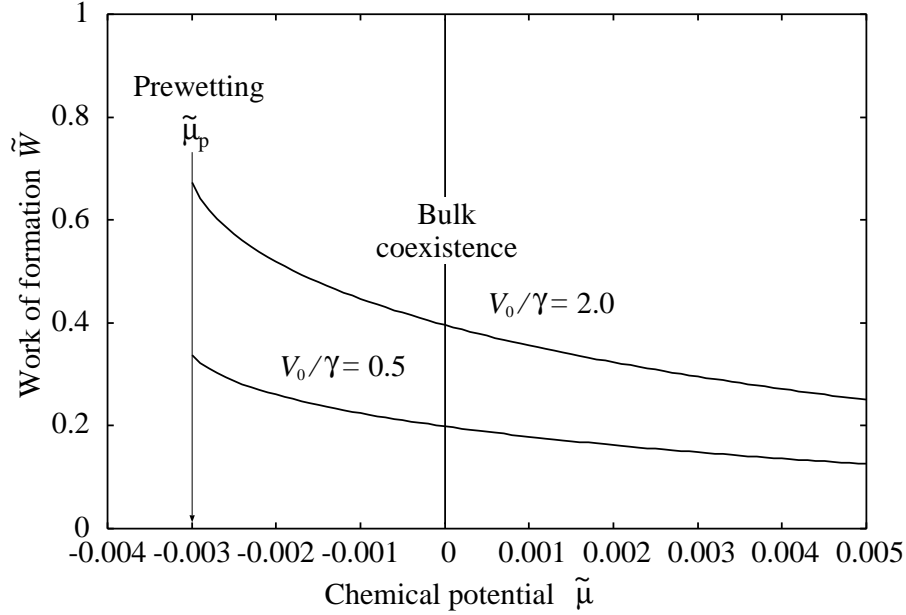}
\end{center}
\caption{
The reduced work of formation $\tilde{W}=W/\gamma$ calculated from Eq.~(\ref{Eq:40x})  in the complete-wetting regime. }
\label{fig:7}
\end{figure}

The reduced work of formation $\tilde{W}=W/\gamma$ from Eq.~(\ref{Eq:40x}) as a function of the reduced chemical potential $\tilde{\mu}$ is shown in Fig.~\ref{fig:7}. In this complete-wetting regime, the work of formation $W$ does not vanish and, therefore, does not agree with the prediction of CNT which suggests $W=0$.  Furthermore, it changes continuously even at the bulk coexistence at $\mu=0$ though the character of the droplet changes from the critical nucleus of the prewetting surface phase transition below bulk coexistence $\mu_{\rm p}<\mu<0$ to the critical nucleus of the heterogeneous bulk phase transition above the bulk coexistence $\mu>0$.  A more detailed discussion on the continuity of the work of formation $W$ as well as that of the derivatives $d^{n}W/d\mu^{n}$ at $\mu=0$ was given in the previous paper~\cite{Iwamatsu2011}

This work of formation $W$ approaches a finite value at the prewetting line $\mu=\mu_{\rm p}$ (Fig.~\ref{fig:7}) for the $d=2$-dimensional droplet as the integral (Eq.~(\ref{Eq:40x})) remains finite.  For such a flat nucleus (Fig.~\ref{fig:6}) with $l_{x}\ll 1$, we can approximate $\left(1+l_{x}^{2}\right)^{1/2}-1\simeq l_{x}^{2}/2$ in Eq.~(\ref{Eq:39x}), and Eq.~(\ref{Eq:40x}) can be approximated by~\cite{Indekeu92}
\begin{equation}
W \simeq \int_{l=l_{e}}^{l=L} \left(2\gamma\Delta\phi\left(l\right)\right)^{1/2} dl.
\label{Eq:42x}
\end{equation}
By using the scaled potential in Eq.~(\ref{Eq:23x}), we observe
\begin{equation}
W \propto \left(\gamma V_{0}\right)^{1/2}l_{0}.
\label{Eq:43x}
\end{equation}
Therefore, the reduced work of formation $\tilde{W}=W/\gamma$ is proportional to the parameter $\sqrt{V_{0}/\gamma}$ in Eq.~(\ref{Eq:33x}).

\subsection*{3.2 $d=3$-dimensional nucleus}
For a $d=3$-dimensional hemispherical droplet, the Euler-Lagrange equation for the free energy functional Eq.~(\ref{Eq:5x}) is given by~\cite{Dobbs99}
\begin{equation}
\gamma \left(\frac{d}{dx}+\frac{1}{x}\right)\left(\frac{l_{x}}{\left(1+l_{x}^{2}\right)^{1/2}}\right)=\frac{dV}{dl}-\mu,
\label{Eq:47x}
\end{equation}
where $l_{x}=dl/dx$ and $x$ is the coordinate measured from the center of the base of the droplet (Fig.~\ref{fig:1}). This equation cannot be integrated as for the $d=2$-dimensional cylindrical droplet. As a result, it is not possible to calculate the height $L$ of the nucleus from Eq.~(\ref{Eq:18x}).  Instead, the height $L$ is determined from the solution of Eq.~(\ref{Eq:47x}) that satisfies the boundary condition $l_{x}=0$ at $x=0$ and $l=l_{e}$ at $x=\infty$.

\begin{figure}[htbp]
\begin{center}
\includegraphics[width=1.0\linewidth]{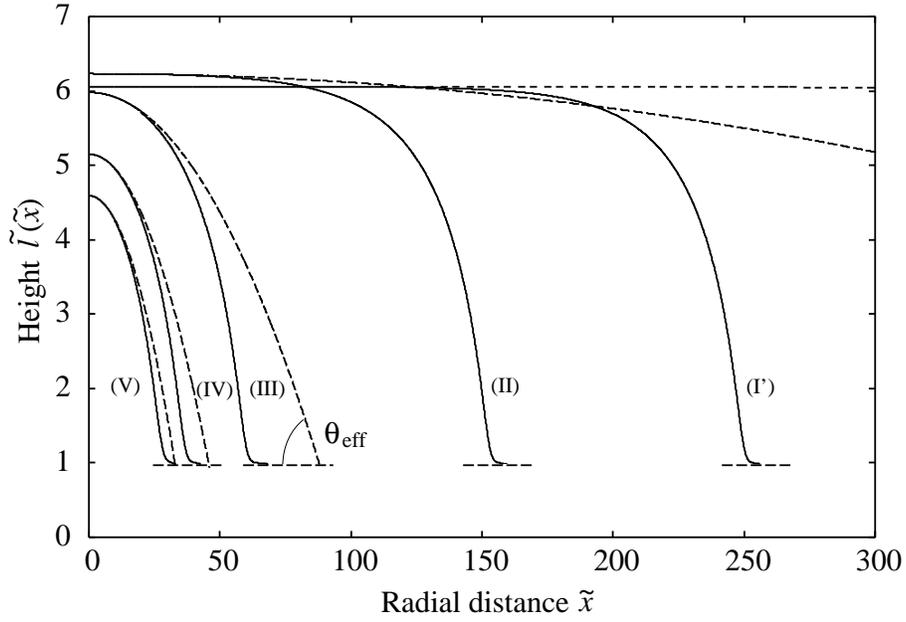}
\end{center}
\caption{
The droplet shape numerically determined from the Euler-Lagrange equation (\ref{Eq:47x}) using the Runge-Kutta method (solid curves) when $V_{0}/\gamma=2.0$ compared with the ideal semi-spherical shape given by Eq.~(\ref{Eq:28x}) (broken curves) for (I') near the prewetting line ($\tilde{\mu}=-0.0027$), (II) below the bulk coexistence and above the prewetting line $0>\mu>\mu_{\rm p}$ ($\tilde{\mu}=-0.0025$), (III) $\tilde{\mu}=-0.0015$, (IV) at the bulk coexistence $\mu=0$ ($\tilde{\mu}=0.0000$), (V) above the bulk coexistence $\mu>0$ ($\tilde{\mu}=+0.0015$).  Only a right half of the droplet is shown.  Note the scales of the vertical and the horizontal axes.} 
\label{fig:8}
\end{figure}

In Fig.~\ref{fig:8} we show numerically determined droplet shapes for various $\mu$ when $V_{0}/\gamma=2.0$ .  In Fig.~\ref{fig:4} we have shown the droplet height $L$ directly determined from the numerically determined droplet shape in Fig.~\ref{fig:8} as the function of the chemical potential $\mu$.  By repeating the argument from Eq.~(\ref{Eq:25x}) to (\ref{Eq:29x}), it is possible to approximate the droplet shape determined from Eq.~(\ref{Eq:47x}) near the top of the droplet by the semi-spherical shape given by Eq.~(\ref{Eq:28x}).  However, the effective radius is now given by
\begin{equation}
R_{\rm eff} = \frac{2\gamma}{\mu_{\rm eff}}
\label{Eq:48x}
\end{equation}
instead of Eq.~(\ref{Eq:29x}).  Then, it is possible to define the effective contact angle $\theta_{\rm eff}$ for a $d=3$-dimensional droplet on a completely-wettable substrate shown in Fig.~\ref{fig:8}. This effective contact angle is also given by Eq.~(\ref{Eq:38x}).

In contrast to the $d=2$-dimensional case, the lateral size of the $d=3$-dimensional droplet is larger and diverges at the prewetting line $\mu=\mu_{\rm p}$. The critical nucleus of the $d=3$-dimensional droplet is more flat.  Using an analogy of classical mechanics, this divergence can be intuitively understandable~\cite{Iwamatsu2011}. 

\begin{figure}[htbp]
\begin{center}
\includegraphics[width=1.0\linewidth]{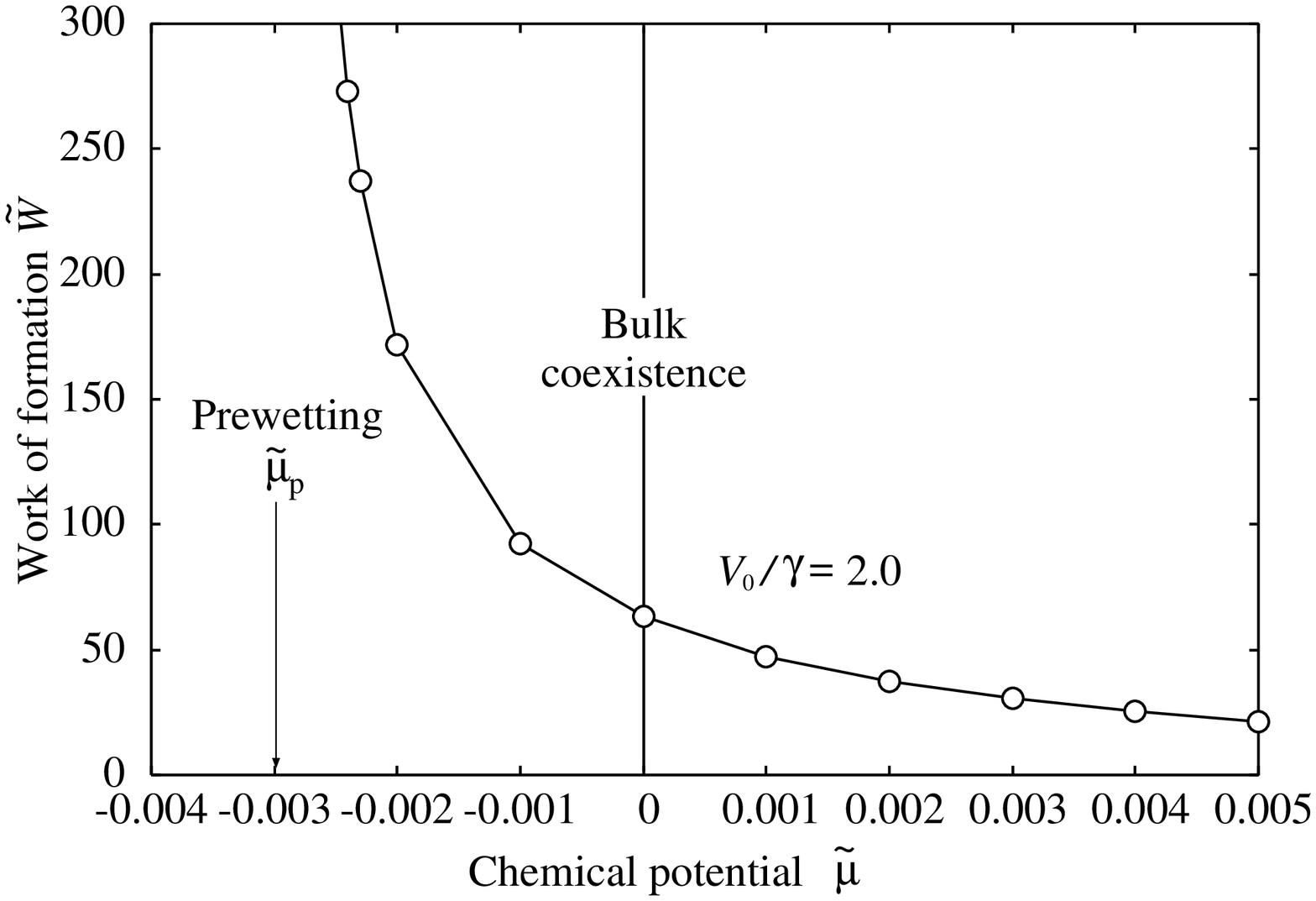}
\end{center}
\caption{
The reduced work of formation $\tilde{W}=W/\gamma$ calculated from Eq.~(\ref{Eq:5x}) using the droplet profile $l(x)$ numerically determined from Eq.~(\ref{Eq:47x}) when $V_{0}/\gamma=2.0$ as a function of the reduced chemical potential $\tilde{\mu}$.  In contrast to the $d=2$-dimensional nucleus, the work of formation $W$ of the $d=3$ dimensional nucleus diverges at the prewetting $\mu_{\rm p}$. }
\label{fig:9}
\end{figure}

Since we cannot integrate Eq.~(\ref{Eq:47x}) analytically for a $d=3$-dimensional droplet, we cannot use Eq.~(\ref{Eq:40x}) to calculate the work of formation $W$.  Instead, we have to resort to a direct numerical integration using Eq.~(\ref{Eq:5x}).  Figure \ref{fig:9} shows the reduced work of formation $\tilde{W}=W/\gamma$ of a $d=3$-dimensional critical nucleus as a function of the reduced chemical potential $\tilde{\mu}$.

Similar to the $d=2$-dimensional critical nucleus, the work of formation $W$ is a continuous function of the chemical potential $\mu$ at the bulk coexistence. However, in contrast to the $d=2$-dimensional nucleus in Fig.~\ref{fig:7}, the work of formation $W$ of the $d=3$-dimensional nucleus diverges at the prewetting line $\mu=\mu_{\rm p}$ in Fig.~\ref{fig:9}.  Indeed the prewetting line acts as an effective bulk coexistence for a completely-wetting substrate.

\section*{\label{sec:sec5}4. Nucleation on a completely-wettable spherical substrate}
In the previous section, it was shown that the nucleation barrier may exist even when the liquid completely wets a flat substrate due to the surface potential which will achieve the prewetting transition.  On a spherical substrate, it is well known that the energy barrier exists even on a completely-wettable substrate since the nucleation is essentially homogeneous nucleation~\cite{Qian2007}.  Since the free energy of a spherical nucleus, which surrounds the spherical substrate with layer thickness $l$ is given by
\begin{eqnarray}
\Omega_{0}\left(l;R_{0}\right)&=&4\pi\left(\gamma\left(R_0+l\right)^2-\frac{1}{3}\mu\left(R_0+l\right)^{3}\right)-4\pi\left(\gamma R_0^2-\frac{1}{3}\mu R_0^{3}\right) \nonumber \\
&=& 4\pi\gamma\left(\left(R^{2}-\frac{2}{3}\frac{R^{3}}{R_{\rm K}^{3}}\right)-\left(R_0^{2}-\frac{2}{3}\frac{R_0^{3}}{R_{\rm K}^{3}}\right)\right),
\label{Eq:49x}
\end{eqnarray}  
where $R_{\rm K}$ is the critical radius of homogeneous nucleation known as the Kelvin radius which is in fact given by the effective radius $R_{\rm K}=R_{\rm eff}$ in Eq.~(\ref{Eq:48x}).  
The energy barrier similar to that of the homogeneous nucleation exists if the radius $R_{0}$ of the spherical substrate is smaller than the Kelvin radius $R_{\rm K}$.

This energy barrier disappears as soon as the radius of the spherical substrate becomes larger than that of the critical nucleus ($R_0>R_{\rm K}$).  Then the barrierless nucleation is achieved~\cite{Qian2007}, which is similar to the barrierless nucleation called athermal nucleation~\cite{Quested2005,Greer2000} on a flat substrate when the radius of the substrate is smaller than the critical radius..  The adsorbed wetting film that surrounds the spherical substrate will grow spontaneously without crossing the energy barrier.  Again the situation that occurred in completely-wettable flat substrate may be achieved. 

Since the surface potential which brings about the first-order wetting transition on a flat substrate will turn barrierless zero-contact angle substrate into a finite-contact angle substrate with finite energy barrier of nucleation, it is expected that a similar behavior will be expected on a spherical substrate even when the radius of the substrate exceeds the critical radius of homogeneous nucleation.  In fact, the effect of surface potential or the disjoining potential on the heterogeneous nucleation on a spherical substrate has been considered by Russian group~\cite{Kuni1996,Kuni2001}.  Since these authors used short-range surface potential and did not pay much attention to the relationship to the surface phase transition, we will extend the work of previous section to explore the possibility of observing the apparent critical nucleus and energy barrier on a completely-wettable spherical substrate when the radius exceeds that of the critical nucleus.

In order to study the effect of surface potential, we must set up the surface potential created by the spherical substrate.  Using the same long-range potential that leads to the model surface potential in Eq.~(\ref{Eq:19x}), we obtain
\begin{equation}
V_{\rm sp}\left(l;R_{0}\right)=V_{0}\left(\frac{1}{2}\left(\frac{l_{0}}{l}\right)^{2}S_{6}\left(z\right) - \frac{1+b}{3}\left(\frac{l_{0}}{l}\right)^{3}S_{7}\left(z\right) + \frac{b}{4}\left(\frac{l_{0}}{l}\right)^{4}S_{8}\left(z\right)\right)
\label{Eq:50x}
\end{equation}
where
\begin{eqnarray}
S_{6}\left(z\right)&=&\frac{2\left(2+4z+3z^2+z^3\right)-z^{2}\left(2+z\right)^{2}\ln\left(\frac{2+z}{z}\right)}{\left(2+z\right)^{2}}, \nonumber \\
S_{7}\left(z\right)&=&\frac{2\left(4+10z+5z^{2}\right)}{\left(2+z\right)^{3}}, 
\label{Eq:51x} \\
S_{8}\left(z\right)&=&\frac{16\left(1+z\right)^{3}}{\left(2+z\right)^{4}}, \nonumber
\end{eqnarray}
are the correction terms due to the finite curvature
\begin{equation}
z = l / R_{0},
\label{Eq:52x}
\end{equation}
of the substrate, where $l$ is the thickness on the wetting film on the spherical substrate with radius $R_{0}$.  In contrast to the previous study~\cite{Bieker1998}, we have used the parameter $z$ defined by Eq.~(\ref{Eq:52x}).  Equation (\ref{Eq:50x}) reduces to Eq.~(\ref{Eq:19x}) since $S_{6}\left(z\right)\rightarrow 1$, $S_{7}\left(z\right)\rightarrow 1$, and $S_{8}\left(z\right)\rightarrow 1$ as $z\rightarrow 0$ ($R_{0}\rightarrow \infty$).

\begin{figure}[htbp]
\begin{center}
\includegraphics[width=1.0\linewidth]{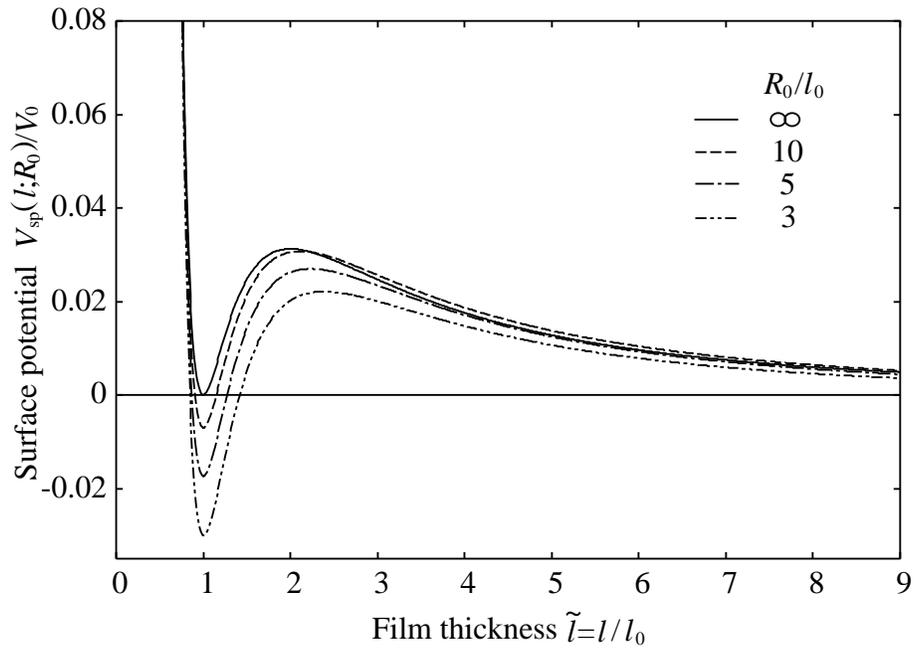}
\end{center}
\caption{
The reduced surface potential Eq.~(\ref{Eq:50x}) when $b=2.0$ that corresponds to the wetting transition point for a flat substrate.  The complete-wetting flat substrate with $R_0\rightarrow\infty$ turns to the incompletely-wetting spherical substrate when the radius $R_{0}$ becomes finite.
 }
\label{fig:10}
\end{figure}

Figure \ref{fig:10} shows the surface potential of a spherical substrate when $b=2$ that corresponds to the surface potential of a flat substrate at the wetting transition point (Fig.~\ref{fig:3-2}).  It can be seen that the substrate turns from complete-incomplete equilibrium (wetting transition point) to incomplete wetting due to the increase of the volume of the spherical interface as the thickness $l$ of the wetting layer increases.  Therefore, a completely-wettable flat substrate may turn to the incompletely-wettable spherical substrate.  The effect of the curvature of the spherical substrate is more pronounced as the radius of the substrate decreases since the incompletely-wetting thin film is more stable.  A similar result was obtained from numerical work using density functional theory~\cite{Napari2003}.  It can be seen from Figs.~\ref{fig:3-2} and \ref{fig:10} that the surface potential created by a spherical substrate is approximated by that created by a flat substrate by renormalizing the parameter $b$ which describes the proximity to wetting transition point.  

Now it is possible to study the effect of surface potential on the free energy of spherical nucleus on a spherical substrate.  The total free energy $\Omega\left(l\right)$ of the nucleus is approximated by
\begin{equation}
\Omega\left(l;R_{0}\right) = \Omega_{0}\left(l;R_0\right) + V\left(l\right),
\label{Eq:53x}
\end{equation}
where $\Omega_{0}\left(l\right)$ is given by Eq.~(\ref{Eq:49x}) and the surface potential $V_{\rm sp}\left(l;R_0\right)$ is approximated by that for the flat substrate in Eq.~(\ref{Eq:19x}) since we are interested in the qualitative effect of the surface potential.

\begin{figure}[htbp]
\begin{center}
\includegraphics[width=1.0\linewidth]{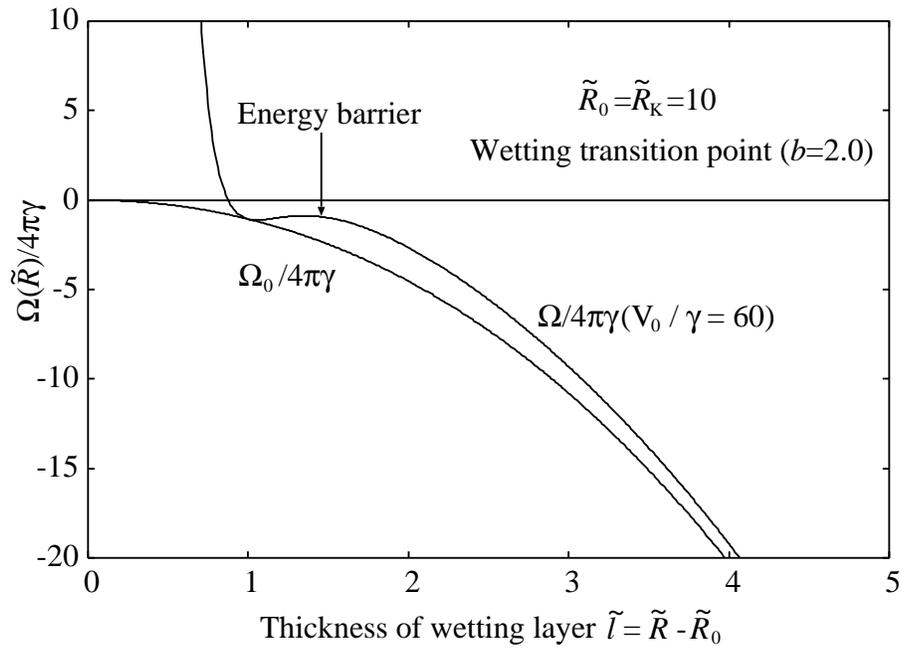}
\end{center}
\caption{
The total free energy given by Eq.~(\ref{Eq:53x}) when $\tilde{R}_{0}=R_{0}/l_{0}=10.0$, $V_{0}/\gamma=60.0$ and $R_{\rm K}=R_{0}$.  In this case with strong surface potential $V_0$, the energy barrier appears due to the strong surface potential which can realize prewetting transitions. 
 }
\label{fig:11}
\end{figure}

Figure \ref{fig:11} shows the full form of the total free energy $\Omega\left(l;R_0\right)$ given by Eq.~(\ref{Eq:53x}) when $\tilde{R}_{0}=R_{0}/l_{0}=10.0$ and the strength of the surface potential is $V_{0}/\gamma=60.0$.  The chemical potential $\mu$ or the Kelvin radius $R_{\rm K}$ is set to $R_{0}=R_{K}$, which is the most favorable condition for the appearance of the energy barrier. The energy barrier appears in Fig.~\ref{fig:11} and, therefore, the critical nucleus with finite contact angle may appear on the completely-wettable spherical surface.  However, the surface potential required to produce energy barrier is fairly-strong.  Also, even if the energy barrier appears, it will not be detectable as long as the energy barrier is lower than the thermal energy $<\sim 30-50kT$.  

\begin{figure}[htbp]
\begin{center}
\includegraphics[width=1.0\linewidth]{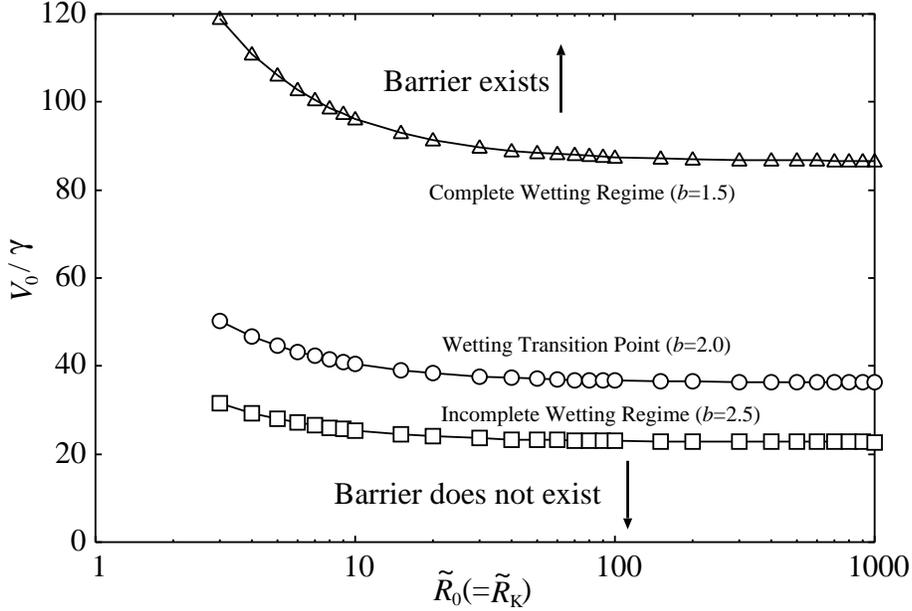}
\end{center}
\caption{
The boundary between barrierless and with barrier nucleation as a function of the radius $R_0$ of the spherical substrate for various surface potentials from incomplete to complete wetting.  Above this boundary with strong surface potential $V_0$, the energy barrier appears in the free energy in Fig.~\ref{fig:11}.
 }
\label{fig:12}
\end{figure}

The boundary for the appearance of the energy barrier is shown in Fig.~\ref{fig:12}.  If the surface potential is larger than this boundary, the energy barrier similar to that in Fig.~\ref{fig:11} will appear.  The potential must be stronger in the completely-wetting regime when $b=1.5$ (see Eq.~\ref{Eq:21x}) than the incompletely-wetting regime ($b=2.0, 2.5$).  Also, the potential must be stronger when the radius $R_{0}$ of the spherical substrate is smaller.  In order to observe the nucleation barrier and critical nucleus on a completely-wettable spherical surface, the surface potential must be much stronger than the typical value of non-polar liquid:
\begin{equation}
\frac{V_{0}}{\gamma}\sim50 \gg 0.1-10,
\label{Eq:54x}
\end{equation}
from Eq.~(\ref{Eq:36x}).  Therefore, the appearance of energy barrier and that of the critical nucleus is less probable on a completely-wettable spherical substrate than on a flat substrate.

\section*{\label{sec:sec4}5. Conclusion}
In this study, we have reviewed our previous work~\cite{Iwamatsu2011} where we used the interface displacement model to study the heterogeneous nucleation on a completely-wettable flat substrate where the surface potential induces the first-order incomplete to complete wetting surface phase transition.  It is pointed out that the classical picture breaks down on a completely-wettable substrate, where CNT predicts that the critical nucleus does not exist and the nucleation barrier is expected to vanish since the apparent contact angle vanishes.  In fact, both the critical nucleus of the heterogeneous bulk phase transition in the oversaturated vapor and the critical nucleus of the surface thin-thick prewetting transition in the undersaturation vapor can exist and transform continuously at the bulk coexistence on a completely-wettable substrate. 

Therefore, the critical nucleus as well as the free-energy barrier of nucleation exist on a completely-wettable flat substrate.  Furthermore, the nucleus exists even under the undersaturated vapor.  The undersaturated vapor turns to an effectively oversaturated vapor due to the effective interface potential of the substrate which prefers infinitely thick wetting films.  The Laplace pressure  (or the effective chemical potential $\mu_{\rm eff}$ in Eq.~(\ref{Eq:29x})) that is necessary to produce semi-circular liquid-vapor interface of the droplet becomes positive by the interface potential {even under the undersaturated vapor}.  Then, the bulk coexistence does not play any role in the critical droplet in the complete-wetting regime.  Instead, the prewetting line plays the role of the coexistence in the complete-wetting regime.  Various properties of the critical nucleus change continuously as functions of the chemical potential.  Therefore, the nucleation rate (Eq.~(\ref{Eq:1x})) is expected to change continuously at the bulk coexistence.  Our results support the conclusion reached from a numerical simulations in Ising system by Sear~\cite{Sear08}. He simply stated that this small nucleus does not "know" whether it will grow to form a wetting layer of finite thickness or a bulk phase of infinite thickness.  

These findings on a flat substrate change drastically on a spherical substrate.  The free energy decreases more rapidly as the thickness of the wetting film increases on a spherical substrate compared to that on a flat substrate because the volume of the film increases more rapidly on a spherical substrate than on a flat substrate. Then, the energy barrier which is necessary to form a critical nucleus of the droplet may disappear.   Therefore, the possibility of observing the activation process of heterogeneous nucleation on a completely-wettable real substrate will depend strongly on the morphology of the substrate.

\section*{acknowledgments}
This work was supported by the Grant-in-Aid for Scientific Research [Contract No.(C)22540422] from Japan Society for the Promotion of Science (JSPS) and the MEXT supported program for the Strategic Research Foundation at Private Universities, 2009-2013.   The author is particularly grateful to Dr. K. L. Mittal for his effort to make the Eighth International Symposium on Contact Angle, Wettability and Adhesion successful and for his hospitality extended to participants during the symposium in Quebec City.  He is also indebted to Professor Ludmila Boinovich for calling his attention to the works conducted by Derjaguin school in Russia.

\end{document}